# Isolation and characterization of few-layer black phosphorus

*Andres Castellanos-Gomez[1,]\*, Leonardo Vicarelli[1], Elsa Prada[2], Joshua O. Island[1], K. L. Narasimha-Acharya[2], Sofya I. Blanter[1], Dirk J. Groenendijk[1], Michele Buscema[1], Gary A. Steele[1], J. V. Alvarez[2], Henny W. Zandbergen[1], J. J. Palacios[2] and Herre S.J. van der Zant[1]*

[1] Kavli Institute of Nanoscience, Delft University of Technology, Lorentzweg 1, 2628 CJ Delft (The Netherlands).

[2] Departamento de Física de la Materia Condensada, Instituto de Ciencia de Materiales Nicolás Cabrera (INC), and Condensed Matter Physics Center (IFIMAC), Universidad Autónoma de Madrid, Cantoblanco, 28049 Madrid (Spain).

a.castellanosgomez@tudelft.nl

ABSTRACT

Isolation and characterization of mechanically exfoliated black phosphorus flakes with a thickness down to two single-layers is presented. A modification of the mechanical exfoliation method, which provides higher yield of atomically thin flakes than conventional mechanical exfoliation, has been developed. We present general guidelines to determine the number of layers using optical microscopy, Raman spectroscopy and transmission electron microscopy in a fast and reliable way. Moreover, we demonstrate that the exfoliated flakes are highly crystalline and that they are stable even in free-standing form through Raman spectroscopy and transmission electron microscopy measurements. A strong thickness dependence of the band structure is found by density functional theory calculations. The exciton binding energy, within an effective mass approximation, is also calculated for different number of layers. Our computational results for the optical gap are consistent with preliminary photoluminescence results on thin flakes. Finally, we study the environmental stability of black phosphorus flakes finding that the flakes are very hydrophilic and that long term exposure to air moisture etches black phosphorus away. Nonetheless, we demonstrate that the aging of the flakes is slow enough to allow fabrication of field-effect transistors with strong ambipolar behavior. Density functional theory calculations also give us insight into the water-induced changes of the structural and electronic properties of black phosphorus.

KEYWORDS

   Introduction

The isolation of single-layer graphene by mechanical exfoliation has unleashed a new research field devoted to the study of the properties of two-dimensional materials [1-3]. In fact, the fabrication of many different 2D materials has been recently demonstrated and mechanical exfoliation has proved to be an effective technique to cleave bulk layered materials down to the single- and few-layer limits. Nonetheless, so far graphene is the only stable elemental 2D material fabricated by mechanical exfoliation; the rest of the exfoliated 2D materials are composed of two or more elements (e.g. BN, $MoS_2$, $Bi_2Se_2$, mica, etc). Although silicene is also an elemental 2D material, [4] it is not stable in free-standing form (it has to be epitaxially grown on specific substrates) and is rapidly destroyed in the presence of low concentrations of oxygen [5]. Recent works have demonstrated that black phosphorus, a layered allotrope of the element phosphorus, can be exfoliated similarly to graphite to fabricate few-layer thick sheets [6-10] triggering the interest in this new elemental 2D material [11-15]. Unlike graphene, few-layer black phosphorus has an intrinsic bandgap, so that field-effect transistors with large current on-off ratios and high mobilities (100-3000 $cm^2$/Vs) can be fabricated [6-10]. Nevertheless, a complete characterization of few-layer black phosphorus flakes is still lacking.

In this paper, we exfoliate black phosphorus to isolate few-layer flakes down to two-layers thick sheets. We present a thorough characterization of the fabricated layers by optical microscopy, atomic force microscopy, Raman spectroscopy, photoluminescence, high-resolution transmission electron microscopy and electronic transport. We show how optical microscopy and Raman spectroscopy can be used to estimate the number of layers of black phosphorus flakes. Transmission electron microscopy measurements also show that exfoliated freely suspended flakes are crystalline and stable. Photoluminescence measurements show spectra that strongly depend on the flake thickness. These results are corroborated by density functional theory (DFT) calculations in combination with exciton binding energy calculations which demonstrate a large dependence of the electronic and optical bandgaps with the number of layers. We also find that black phosporous is hydrophilic: water molecules easily stick to it, presenting large adsorption binding energies which can be an asset in sensing or catalysis applications.

**Sample fabrication**

Black phosphorus is the most stable allotrope of phosphorus, which also occurs in the form of white, red and violet phosphorus. In its bulk form, black phosphorus is a direct-gap semiconductor with a 0.33 eV bandgap and mobilities of up to 20000 cm$^2$/V·s at room temperature [16-18]. Unlike other allotropes, black phosphorus is characterized by a layered structure: the in-plane bonds are strong and the van der Waals interlayer interaction is weak[19]. The crystal structure of bulk black phosphorus is orthorhombic, with space group *Cmca*. Note that films with a discrete number of layers there is no translational symmetry along the *z* axis (out of plane) and thus this configuration cannot be described with space group *Cmca*. Figure 1 shows a representation of the black phosphorus crystalline structure formed by a puckered honeycomb lattice. Note, that all the lattice parameters and angles displayed in the figure have been obtained through *ab-initio* calculations of the crystal structure (see band structure calculation section for more details). This layered structure permits to employ mechanical exfoliation to extract thin black phosphorus from a bulk crystal. Nonetheless, we found that the mechanical exfoliation protocol to obtain graphene or transition metal dichalcogenides typically yields a very low density of atomically thin black phosphorus flakes.

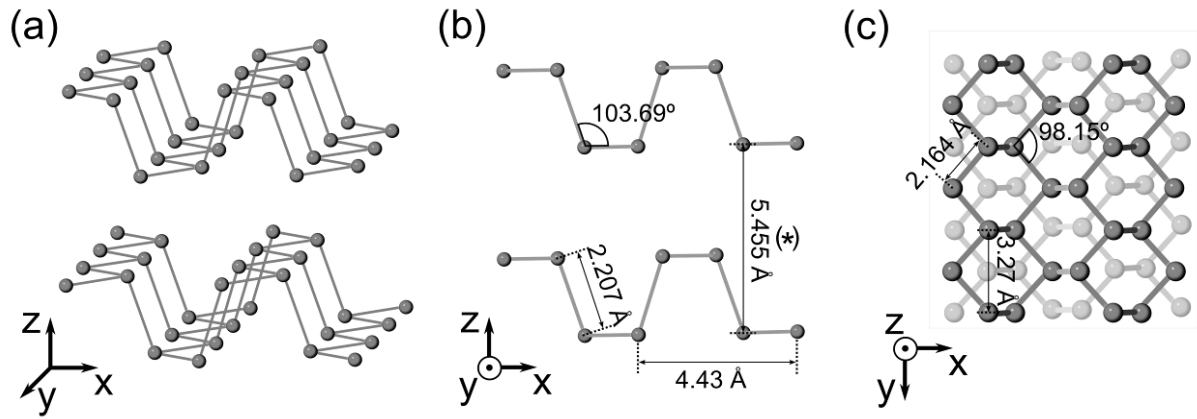

**Figure 1. Black phosphorus structure.** Schematic diagram of the crystalline structure of black phosphorus obtained from the relaxation of the structure using density functional theory calculations (see last part of the manuscript). The layered structure is composed of sheets with the phosphorus atoms arranged in a puckered honeycomb lattice. Adjacent layers interact by weak van der Waals forces and are stacked following an ABA stacking order. (a) 3D representation. (b) Lateral view. (c) Top view. The distances and angles displayed have been obtained by relaxing the crystal structure of a single-layer black phosphorus sheet. The distance value marked with (*) has been obtained by relaxing the bulk structure.

We have modified the mechanical exfoliation technique to optimize the deposition of atomically thin black phosphorus flakes. We found that conventional mechanical exfoliation with adhesive tape yields a low density of few-layer black phosphorus flakes and leaves traces of adhesive on the surface. Employing an intermediate viscoelastic surface to exfoliate the flakes substantially increases the yield and reduces the contamination of the fabricated flakes. In the following, the fabrication steps are described. A piece of commercially available bulk black phosphorus (99.998%, Smart Elements) is cleaved several times using blue Nitto tape. The tape containing the thin black phosphorus crystallites is then slightly pressed against a poly-dimethilsiloxane (PDMS) based substrate and peeled off rapidly. Transmission optical microscopy is used to identify thin flakes and to reliably distinguish them from thick black phosphorus. The thin flakes on the surface of the PDMS substrate can be transferred to other substrates by simply putting the PDMS substrate in gentle contact with the new acceptor substrate and peeling it off slowly (it takes about 5-10 minutes to peel off completely the stamp from the surface). The process is based on a recently developed all-dry transfer technique and we refer the readers to Ref. [20] for more details. A comparison between the results of employing conventional mechanical exfoliation and the PDMS-based method described here can be found in the Supporting Information.

Figure 2a shows a transmission mode optical image of a black phosphorus flake deposited by mechanical exfoliation onto a PDMS substrate. We found that the absorbance of black phosphorus flakes is a multiple of 2.8 % (an analysis of the optical transmittance of black phosphorus on PDMS substrates can be found in the Supporting Information). We thus infer that 2.8 % is the absorbance of a single-layer. The thinner part of the flake shown in Figure 2a has an optical absorbance of 5.5 ± 0.2 %, which corresponds to a bilayer black phosphorus flake. The flake shown in Figure 2a has been transferred onto a silicon chip with 285 nm of thermally grown $SiO_2$ (Figure 2b). Note that part of the flake broke during the transfer. Figure 2c shows an atomic force microscopy (AFM) topographic image acquired in the region highlighted with a black square in Figure 2b showing that the transferred black phosphorus flake is 1.6 nm thick. This topographic height corresponds to a thickness of three black phosphorus layers. Nonetheless, one has to take into account that an interfacial layer of adsorbates might artificially increase the height determined by AFM. In the case of single-layer reduced graphene oxide, for instance, an AFM height of 1 nm is often measured while one would expect a height value of ~0.35 nm [21].

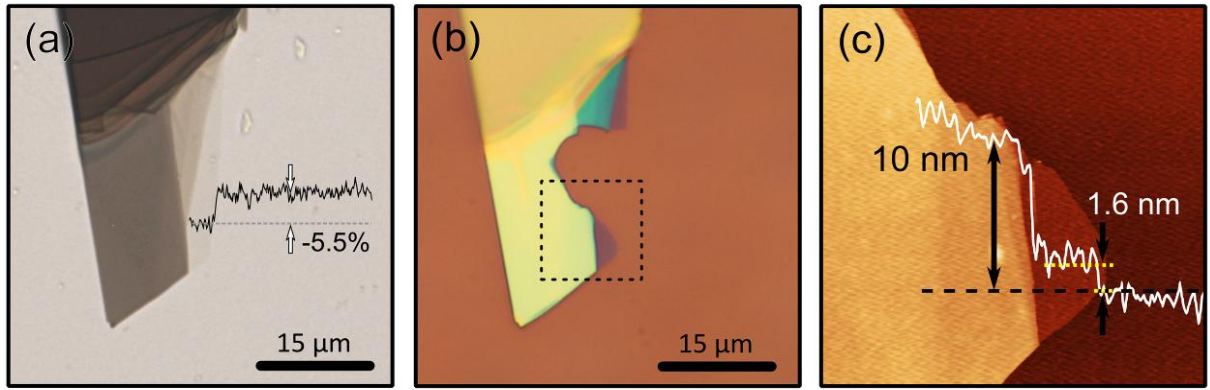

**Figure 2. Isolation of few-layer black phosphorus flakes.** (a) Transmission mode optical microscopy image of a few-layer black phosphorus flake exfoliated onto a PDMS substrate. An optical transmittance line profile has been included in the figure to highlight the reduction of about 5.5% in the optical transmittance in the thinner part of the flake. (b) Bright field optical microscopy image of the same flake after transferring it onto a SiO2/Si substrate. Note that part of the flake wan broken during the transfer. (c) Atomic force microscopy topography image of the region highlighted with a dashed square in (b). A topographic line profile, acquired along the horizontal dashed black line, has been included in the image.

**Raman spectroscopy**

Raman spectroscopy is a very powerful tool to characterize 2D materials [22]. Here, we present Raman spectroscopy characterization of black phosphorus flakes with different thicknesses. Figure 3b shows the measured Raman spectra for the same black phosphorus flake with thicknesses ranging from 1.6 nm to 9 nm. In the spectra there are four prominent peaks at 362.1 cm$^{-1}$, 439.5 cm$^{-1}$, 467.7 cm$^{-1}$ and 520.9 cm$^{-1}$. The peak at 520.9 cm$^{-1}$ corresponds to the Raman peak of the silicon substrate. The peaks at 362.5 cm$^{-1}$, 439.8 cm$^{-1}$ and 467.1 cm$^{-1}$, on the other hand, are due to vibrations of the crystalline lattice of the black phosphorus and they match the Raman shifts attributed to the $A^1_g$, $B_{2g}$ and $A^2_g$ phonon modes observed in bulk black phosphorus [23]. This indicates that exfoliated black phosphorus flakes remain crystalline after the exfoliation. Figure 3a shows a schematic of the black phosphorus lattice, indicating the vibration direction of the phosphorus atoms in the different Raman modes [23]. While the $B_{2g}$ and $A^2_g$ modes correspond to vibrational modes where the atoms oscillate within the layer plane, in the $A^1_g$ mode the phosphorus atoms vibrate out-of-plane.

We found that the position of the Raman peak positions slightly depend on the number of layers of the black phosphorus flakes (see Supporting Information). However, the shift is not pronounced enough to provide an accurate method to determine the number of layers with the resolution of our Raman system (~0.5 cm$^{-1}$). The intensity ratio between the $A^1_g$ peak and the silicon peak, on the other hand, can be used to determine the thickness within 1-2 nm accuracy (see Figure 3c). Note that the intensity of the silicon peak depends on its crystalline orientation with respect to the laser polarization and thus the substrate has to be carefully aligned to allow accurate sample-to-sample comparisons. The ratio between the intensity of the $B_{2g}$ (or $A^2_g$) and silicon peaks strongly varies from flake to flake. This variation might be due to different crystalline orientations of the flakes with respect to the laser, as the $B_{2g}$ and $A^2_g$ modes are due to vibrations within the layer plane (see the Supporting Information).

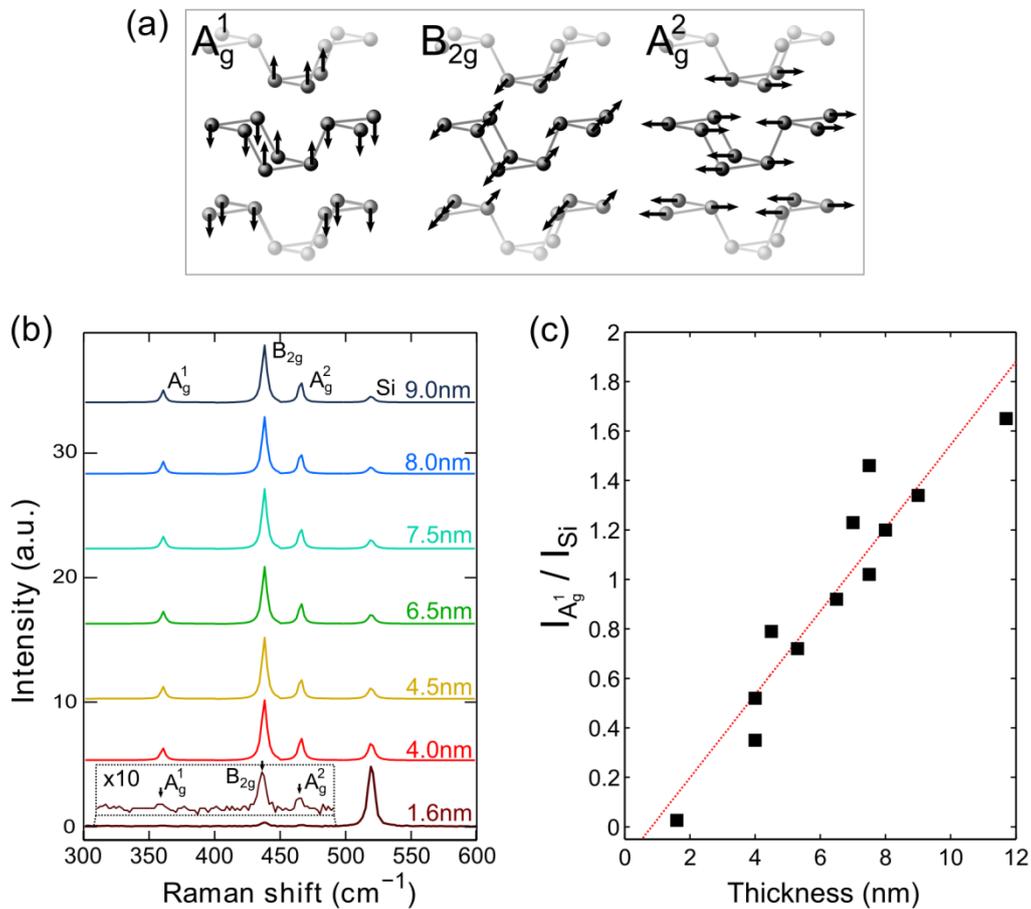

**Figure 3. Raman spectroscopy of fabricated black phosphorus flakes.** (a) Schematic of the different Raman-active vibrational modes of the black phosphorus lattice. (b) Raman spectra measured with a 514 nm excitation laser in different zones of a black phosphorus flake with thicknesses ranging from 9 nm down to 1.6 nm. (c) Thickness dependence of the intensity ratio between the $A^1_g$ and Si peaks.

**Transmission electron microscopy**

We employed transmission electron microscopy (TEM) to further characterize the crystallinity of the exfoliated black phosphorus flakes. In order to make possible to use high-resolution transmission electron microscopy (HRTEM), to allow for direct imaging of the atomic structure of the sample, the studied flakes must be freely-suspended. Here we exploit the fact that our fabrication method allows one to transfer the atomically thin black phosphorus flakes onto different substrates, making possible to deposit thin black phosphorus flakes onto silicon nitride membranes with holes [20]. Figure 4a shows an optical microscopy image of the deposited flake, the thinner part has an optical absorbance of 5.8 ± 0.4 % with respect to the $Si_3N_4$ membrane which corresponds to a bilayer.

The sample has been loaded in a transmission electron microscope FEI Titan right after the transfer to avoid sample contamination. The TEM imaging has been carried out at an acceleration voltage of 300 kV. An HRTEM image from a multilayer area of the sample is shown in Figure 4b. The uniformity in this image indicates that the lattice contains no extended defects (single vacancies cannot be detected). Therefore, few-layer black phosphorus flakes are stable and crystalline even in free-standing form. Since very thin areas were observed to be very beam sensitive, we chose a very low beam intensity and Electron Diffraction (ED) with a large illumination area of 400 nm in diameter to study their crystal structure. No serious amorphisation of the thin sample was observed. ED patterns were recorded with 0° tilt angle at various locations of the flake.

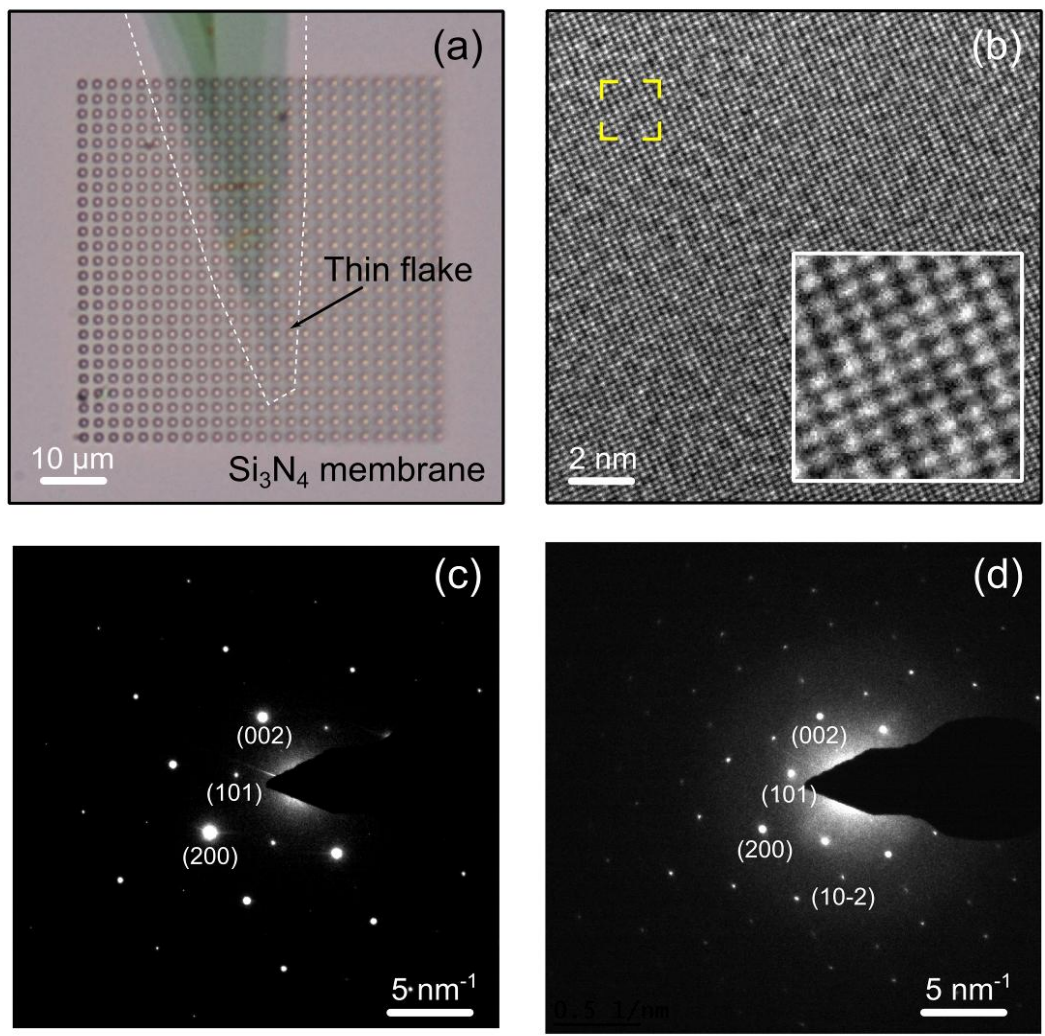

**Figure 4. Transmission electron microscopy study of few-layer black phosphorus flakes.** (a) Optical image of a black phosphorus flake transferred onto a holey silicon nitride membrane. (b) High resolution transmission electron microscopy image of the multilayered region of the flake (~ 13-21 layers). (c) and (d) are electron diffraction patterns acquired with a 400 μm spot on the thick (~ 13-21 layers) and the thin (~2 layers) region of the flake.

Interestingly, we found that electron diffraction patterns depend on the number of layers and thus ED can be employed to determine the thickness of the black phosphorus flakes. We simulated electron diffraction patterns finding that the ratio between the 101 and 200 reflections depends on the number of black phosphorus layers as indicated in Table 1. Note that the use of the I(101)/I(200) ratio to determine the layer number is only valid if the diffraction pattern is taken from an area with uniform thickness.

Figure 4c and Figure 4d show the diffraction pattern corresponding to a thick region and a thin region of the flake, respectively. In the ED pattern of the thick region we notice strong $h,l = 2n$ reflections and weak $h,l = 2n+1$ reflections. This is consistent with our calculations, as the I(101)/I(200) ratio should be very small for thick black phosphorus flakes if the number is uneven and zero if the number is even. Based on this ratio, we estimated that the thick sample corresponds roughly to 21 layers. A refinement using MSLS software [24] of this diffraction pattern, using only the reflections with $h,l = 2n$, indicates the thickness to be 7 nanometers (13 layers circa) with an R-value of 0.001 %.

| Number of layers | I(101)/I(200) |
|---|---|
| 1 | 2.557 |
| 2 | 0.001 |
| 3 | 0.286 |
| 4 | 0.001 |
| 5 | 0.104 |
| 6 | 0.001 |
| 21 | 0.009 |
| Exp. data thin flake 1 | 0.31 |
| Exp. data thin flake 2 | 0.42 |
| Exp. data thick flake | 0.01 |

**Table 1. Thickness dependence of the electron diffraction patterns.** We display the thickness dependence of the intensity ratio between the 101 and 200 reflections. The experimental data acquired on two spots of the thin flake and one spot of the thicker area has been included for comparison.

In the ED pattern of the thin region we notice that the sum of the intensities of the $h,l = 2n$ reflections is almost equal to that of the $h,l=2n+1$ reflections. In particular, we measured

I(101)/I(200) ratios of 0.29 and 0.42 in two separate positions of the thin region. According to our calculations (see Table 1), these values fall in the range corresponding to a thickness of 1 to 3 layers. This ratio does not match with a specific value of a single type of layer, and therefore we expect that the thickness of the sample is non-uniform in the illuminated area (circle with 400 nm diameter). For example, an area with 25% of monolayer and 75% of double layer has an the I(101)/I(200) ratio of 0.36. Note that this thickness estimation agrees fairly well with the one obtained from the optical absorbance analysis. Another interesting feature is the presence of "forbidden" reflections $h+l = 2n+1$ in the thin sample. We measured that these reflections account for 5% of the total intensity of the diffracted beams, almost 10 times more than the value expected from the calculations for a monolayer (see Supplementary Information). The presence of these strong forbidden reflections might be explained by the presence of adatoms on the surface of the black phosphorus layer or a slight distortion of the lattice.

**Electronic properties of exfoliated black phosphorus flakes**

*Photoluminescence measurements*

Photoluminescence measurements have been previously employed to characterize the optoelectronic properties of 2D semiconductors, such as the transition dichalcogenide $MoS_2$, finding a thickness dependent band structure due to the vertical confinement of the charge carriers when the thickness is reduced [25, 26]. We study the black phosphorus flakes by photoexciting them with a green laser (with 2.41 eV) and measuring the energy of the emitted photons. Figure 5 shows a comparison between the spectra measured on black phosphorus flakes with different thicknesses, ranging from 1.6 nm to 5.2 nm. While the thinner flakes (~1.6 nm, bilayer flakes according to their optical contrast) present a marked peak sticking out of the background, the spectra of thicker flakes remain featureless. The lack of a photoluminescence peak in the thick black phosphorus flakes is expected as bulk black phosphorus is a semiconductor with a direct bandgap of about 0.33 eV, which is outside the measurement window. However, the peak observed for the thinner flakes indicates that these flakes exhibit a larger optical bandgap of about 1.6 eV. We rule out the laser-induced amorphization or chemical modification as the origin of the observed photoluminescence peak as amorphous phosphorus shows a photoluminescence at smaller energies (1.18-1.36 eV) [27, 28] and strongly irradiated black phosphorus develops a photoluminescence peak at 2.0-2.1 eV probably due to laser-induced chemical modification (see the Supporting Information). Nevertheless, we found that during the photoluminescence measurements the thinner black phosphorus region became destroyed, probably due to laser-induced oxidation (see Supporting Information), which might affect the reliability of the bandgap estimation obtained by photoluminescence measurements. Further studies under controlled atmospheres are needed to accurately determine the thickness dependence of the photoluminescence peaks. Note that the laser-induced damage of the thin black phosphorus flakes limited our integration time yielding weak photoluminescence spectra but the amplitude of the photoluminescence peak is larger than that of the Raman peaks (see Supporting Information) similarly to what is observed in other direct bandgap 2D semiconductors [25, 26].

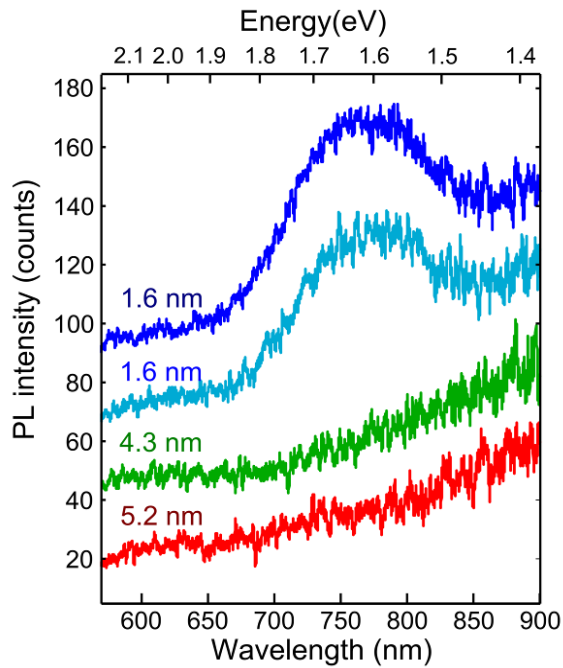

**Figure 5. Photoluminescence measurements on exfoliated black phosphorus flakes.** Photoluminescence spectra acquired on black phosphorus flakes with different thicknesses (indicated in the plot). The spectra measured on thin flakes (1.6 nm thick, 2 layers according to the optical transmittance) show a clear peak around 1.6 eV while the spectra measured on thicker flakes are featureless within the measurement window. Note that the thinner flakes were destroyed during the photoluminescence measurement by laser-induced damage, probably by laser-induced oxidation.

*Band structure calculations*

The photoluminescence measurements indicate a thickness dependent band structure of the black phosphorus flakes, calling for a calculation of the band structure as a function of the number of layers. We have performed density functional theory (DFT) calculations of the atomic and electronic structure of black phosphorus for bulk, multilayer, and monolayer cases. Due to the weak interlayer coupling an appropriate treatment of van der Waals interactions, particularly in the bulk and multilayer cases, is required [29, 30]. The resulting atomic structure is shown in Figure 1. For the calculation of the electronic structure, however, we use the CRYSTAL code [31] because of its implementation of the B3LYP functional which includes a portion of Hartree-Fock (HF) exchange which helps restoring the well-known band gap deficiency of standard DFT. Figure 6 shows the evolution of the calculated band structure with the number of layers. The obtained value for the gap in bulk (1.08 eV) is certainly larger than the experimental one of 0.33 eV [18] and the one obtained from standard (non-hybrid) DFT calculations (see Table 2 and Refs. [32, 12]). This is not a surprise since HF exchange does not work well for metals or small gap semiconductors, overestimating the gaps in the latter. However, as the number of layers decreases and the band gap strongly increases we expect B3LYP functional to perform better. In fact, the band gap we obtain for another well-known two-dimensional crystal such as $MoS_2$ is 2.85 eV, which lies in the range of values reported in the literature [33] and give us confidence in the choice of the B3LYP functional for the calculation of the electronic structure of the monolayer and, possibly, of few-layer systems. For comparison we include in Table 2 the values of the gap obtained using a standard GGA functional. Now the gaps are certainly smaller than those obtained with the B3LYP functional. Although the GGA functional seems to perform better for the bulk case, the gap values obtained for few layers are clearly below the experimental PL peaks even without including excitonic effects (see below). Leaving aside the bandgap discussion, the calculated band structure is similar to others reported in the literature [32, 12, 11, 6, 34]. In particular, highly anisotropic electron and hole masses are obtained [35]. These are shown in Table 1 and used for the exciton binding energy in the next section.

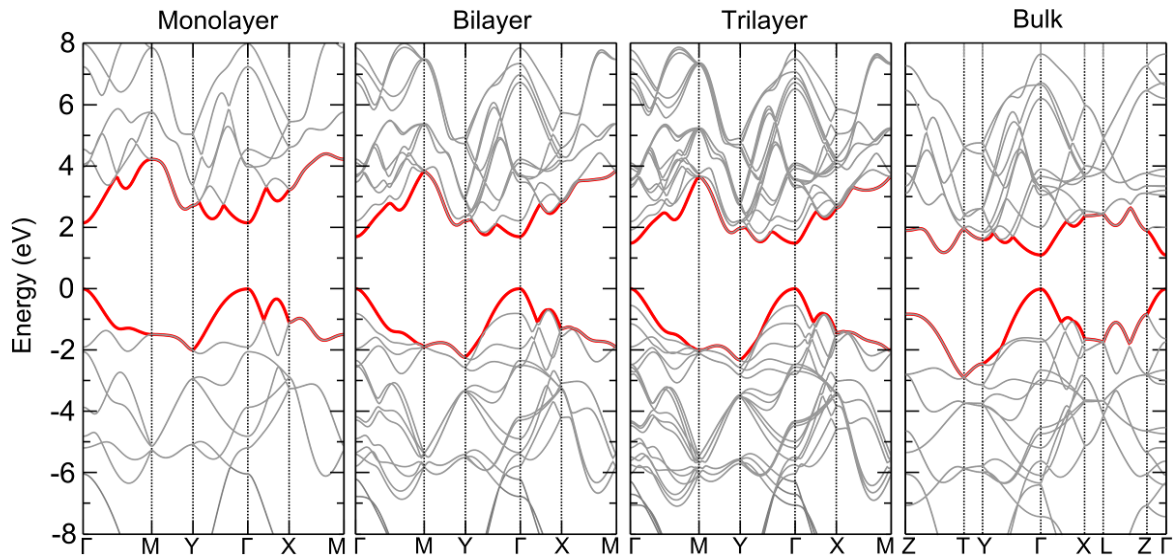

**Figure 6. Calculated band structure.** Calculated electronic band structure for monolayer, bilayer, trilayer and bulk black phosphorus sheets at all high-symmetry points in the Brillouin zone. The energy is scaled with respect to the Fermi energy $E_F$.

*Exciton binding energy of black phosphorus*

The bandgap determined by photoluminescence spectroscopy (usually referred to as the optical bandgap) is lower than the electronic bandgap as the generated electron/hole pairs (excitons) are bound by the Coulomb interaction which reduces the energy of the excitons with respect to free electron/holes. The energy needed to separate the electron/hole pairs, the exciton binding energy, is the difference between the electronic bandgap and the optical bandgap [36, 37]. Therefore in order to compare our photoluminescence results with the band structure obtained with DFT (B3LYP), one has to account for the exciton binding energy.

We estimate the exciton binding energy $E_b$ of black phosphorus layers employing the Wannier effective-mass theory [38] in two-dimensions. We accounted for the difference in dielectric constants between the air and substrate and for the anisotropic effective masses of the electrons and holes (see Materials and Methods Section and Supporting Information for more details). Table 2 summarizes the results of the binding energy calculations for black phosphorus flakes over a $SiO_2$ substrate with different number of layers. The optical bandgap can be readily obtained by subtracting the exciton binding energy value from the electronic bandgap determined through the DFT calculations. For instance, the optical bandgap for a single-, bi- and tri-layer black phosphorus flakes would be 1.77 eV, 1.45 eV and 1.29 eV respectively. Note the fairly good agreement with the 1.6 eV value of optical bandgap obtained for bilayer flakes from photoluminescence measurements. The calculated values also show that for thicker flakes the optical bandgap is too small to be detected with our photoluminescence setup, explaining the featureless spectra measured for multilayered flakes.

Note that, recently, the optical spectra of monolayer black-phosphorus in vacuum have been calculated using first-principles GW-Bethe-Salpeter equation [14]. The reported exciton binding energy is around 0.8 meV. Using our variational approach for the same case of a monolayer in vacuum, we obtain 0.71 eV, which closely agrees with this sophisticated methodology. Actually, the use of effective-mass models to treat excitonic effects is widespread in the literature of transition metal dichalcogenides [39-41], giving accurate results in addition to the possibility of straightforwardly treating different dielectric environments.

| # of layers | B3LYP electronic bandgap (eV) | GGA electronic bandgap (eV) | $\mu_x$ (m$_e$) | $\mu_y$ (m$_e$) | $E_b$ (eV) | $\lambda_x$ (Å) | $\lambda_y$ (Å) |
|---|---|---|---|---|---|---|---|
| 1 | 2.15 | 1.07 | 0.089 | 0.650 | 0.38 | 7.63 | 3.78 |
| 2 | 1.70 | 0.70 | 0.097 | 0.576 | 0.25 | 9.79 | 5.24 |
| 3 | 1.48 | 0.52 | 0.102 | 0.528 | 0.19 | 11.37 | 6.40 |
| 4 | 1.36 | 0.43 | 0.100 | 0.506 | 0.16 | 7.46 | 13.07 |

**Table 2. Summary of the calculated bandgaps and exciton binding energies.** We display the thickness dependence of the electronic bandgap, effective masses, exciton binding energies over a SiO$_2$ substrate and exciton radii in the x and y directions.

**Environmental stability of the exfoliated flakes**

We also characterized the stability in time of exfoliated flakes. We found that (when kept in ambient conditions) quickly after sample fabrication droplets appear on the surface of black phosphorus flakes. Figure 7a shows a temporal sequence of optical images of a black phosphorus flake after its transfer. After one hour, droplets become visible on the surface and keep growing (see fourth panel in Figure 7a). Note, that after storing the sample in a vacuum chamber for a few hours these droplets disappear. We additionally found that a long exposure to air (more than one week) deteriorates the black phosphorus etching away the thinner parts of the flakes (Figure 7b-c). Nonetheless, the flakes are stable enough to fabricate electronic devices [8, 7, 6] as the nanofabrication procedures can be carried out in about 4 hours, keeping the samples in vacuum between each processing step. Figure 8 shows the Raman spectroscopy characterization of the sample shown in Figure 7c after its long exposure to air. While the Raman spectra acquired at different locations on the thin part of the flake do not show any Raman peak associated with black phosphorus, the spectra acquired on the thicker part of the flake still shows the characteristic peaks expected for black phosphorus. This indicates that the thick part of the flake has not been fully etched away nor chemically modified after its long exposure to air.

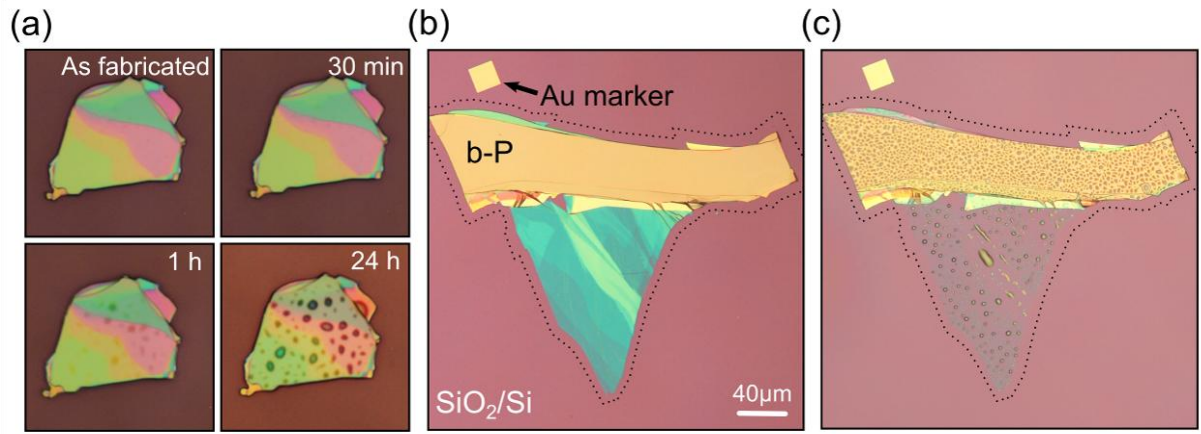

**Figure 7. Aging of black phosphorus flakes.** (a) Sequence of optical images acquired at different times after the transfer of the exfoliated black phosphorus flakes. The sequence shows how one hour after the transfer some droplet-like structures become visible on the surface of the flakes and how they keep growing when the samples are kept in air. (b) and (c) shows the comparison of a sample right after the transfer and after two weeks in air, respectively.

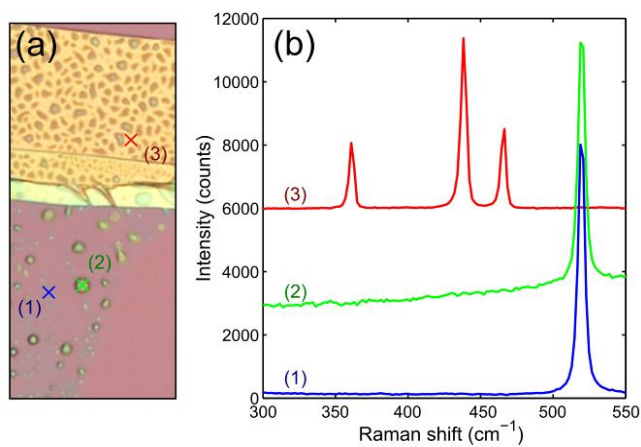

**Figure 8. Raman spectroscopy characterization of aged flakes.** (a) Zoomed-in image of the flake shown in Figure 7c. (b) Raman spectra acquired on the regions indicated by crosses and numbers in (a).

We attribute the presence of droplets on the surface of the exfoliated flakes (Figure 7a) to adsorbed water. In fact, previous theoretical work showed that black phosphorus presents a strong dipolar moment out of plane which makes it very hydrophilic [32]. We performed a micro-droplet condensation experiment [42] to prove whether the appearance of droplets on the surface is due to the condensation of air moisture or not. Black phosphorus flakes, transferred onto a $SiO_2$/Si surface, are inspected under an optical microscope while water is condensed by simply exhaling it through a straw over the sample. Figure 9 shows a sequence of video frames of the micro condensation process. The microscopic droplets start to condensate at the same time on the substrate and on the surface of the black phosphorus. When the humidity is further increased, the droplets on the surface of the flake grow and coalesce. From Figure 9d one can see that the shape of the droplets on the flake is not spherical as they are wetting the flake, demonstrating the high hydrophilic character of the surface.

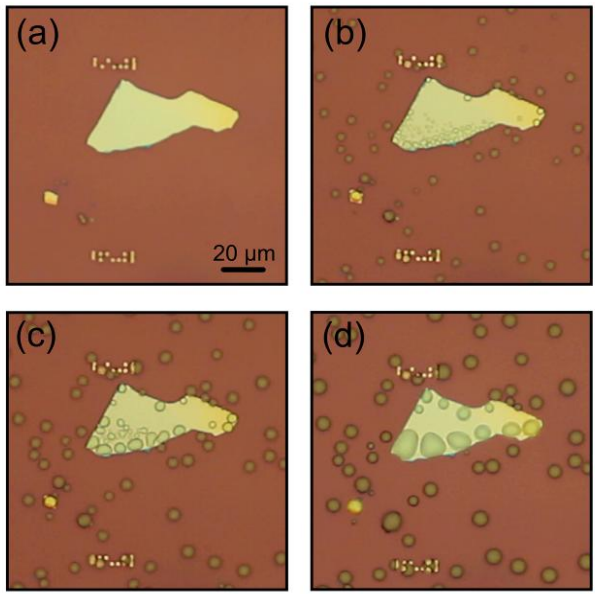

**Figure 9. Water condensation on black phosphorus flakes.** (a) to (d) Sequence of optical images acquired during a microdroplet condensation experiment. The humidity is increased from (a) to (d).

We have verified the hydrophilic character of black phosphorus by performing DFT calculations with the SIESTA code [29]. We place water molecules on the surface of black phosphorus and relax the atomic structure until stable configurations are found. Figure 10a shows an example with one water molecule deposited on a free-standing monolayer. The supercell contains 26 atoms which corresponds to a water concentration of $1.7 \cdot 10^{13}$ cm$^{-2}$. Water and black phosphorus present strong dipoles [32] and, as expected from dipole-dipole interactions, we observe a large distortion of the lattice structure, which shrinks by around 25%. The band gap, in turn, changes also by 20%. Figure 10b shows an example with three molecules adsorbed on a bilayer system. The molecules stick to the topmost layer on different adsorption sites, but always with the O closer to the surface. We also observe that in bilayer structures the topmost layer gets distorted upon adsorption of the water molecules while the bottom one essentially retains its crystallographic structure. This bottom layers helps stabilizing the upper one which becomes less distorted than a free-standing monolayer.

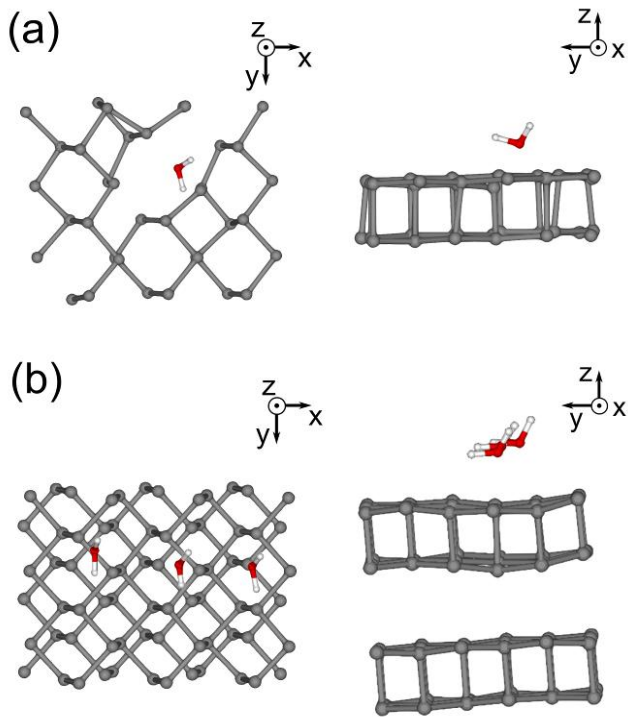

**Figure 10. Calculated water adsorption on black phosphorus.** (a) Schematic top and side view of the atomic structure of a monolayer of black phosphorus with water molecules adsorbed as obtained from DFT calculations. (b) The same but for a bilayer with a higher concentration of water molecules (three per supercell here). The upper layer gets more distorted than the bottom one, but less than the monolayer in panel (a).

This highly hydrophilic behavior found in black phosphorus flakes differentiates it from other 2D materials such as graphene or $MoS_2$ that show a marked hydrophobic behavior. A strong hydrophilic character might be a key strength of black phosphorus in certain applications such as support for biomolecules or sensors, where hydrophobicity of other 2D materials is a drawback [43, 44]. For instance the hydrophobic properties of pristine graphene prevent the direct adsorption of many bio-molecules hampering its use as bio-sensor. Furthermore, the non-specific hydrophobic interactions between DNA molecules and pristine graphene also limits its use as nanopore material in DNA sequencing because of the rapid clogging of the nanopores [44, 45]. Therefore, black phosphorus flakes can find applications in the future where the combination between its electronic and hydrophilic properties are exploited.

**Field effect devices based on few-layer black phosphorus flakes**

As pointed out in the previous section, despite of the aging of the black phosphorus flakes (when kept in air) one can employ them to fabricate electronic devices as the typical nanofabrication processes can be accomplished in about 4 hours. Note also that during most of the processes the flakes are kept in vacuum conditions (as during electron beam lithography and metal evaporation) and/or capped with a polymer layer which helps to preserve the flakes minimizing its exposure to air.

Figure 11a shows an optical image of a field-effect device with two channels lengths (450 nm and 150 nm), fabricated from a 12 nm thick black phosphorus flake. Figure 11b and Figure 11c show the conductance ($G$) *vs*. back-gate voltage ($V_g$) transfer characteristics measured for the longer and shorter channel, respectively. The device exhibits ambipolar transport with a slight asymmetry between holes and electrons. Note that most of the reported devices in the literature show a strong p-type behavior and a rather poor electron conduction which could be due to the difference in the contact metal employed in our devices [6-10]. Using the $G$-$V_g$ data of this device we calculate the transconductance ($\partial G / \partial V_g$) to estimate the mobility from

$$\mu = \frac{L}{WC_{\text{ox}}} \cdot \frac{\partial G}{\partial V_{\text{g}}} \tag{1}$$

where $L$ is the channel length, $W$ is the channel width, and $C_{\text{ox}}$ is the oxide capacitance per unit area calculated using a parallel plate capacitance model [46]. In Table 3 we summarize the room temperature and low temperature mobilities for each channel. Note that the mobility values are lower bound estimates as the contact resistance is included in the two terminal measurements. Overall, we measure higher mobilities for the hole carriers as compared with electrons which is in agreement with recent predictions [11]. Carrier mobilities are lower for the shorter channel length. This is consistent with reports on MoS$_2$ transistors and silicon devices with short channels where the contact resistance has an stronger effect in the total resistance of the devices and the carrier velocity saturates resulting in lower estimated mobilities [47]. The hole mobility of 35 cm$^2$/Vs measured for the 450 nm channel device is a factor 5-10 lower than other black phosphorus field effect devices [6-9]. Further work needs to be done to determine if this reduced mobility value is due to the role of the Al contacts employed (instead of Au used in previous works) or is due to the short channel length of our devices (shorter than previous devices). At low temperature (4K) the mobility shows a moderate decrease (~10%) and the threshold voltages for holes and electrons increase as the thermally activated carriers are removed and the on/off ratios become larger.

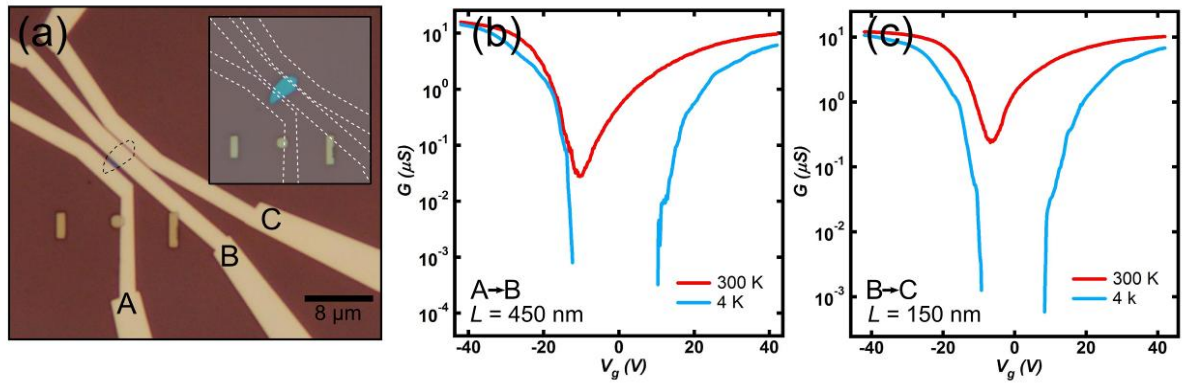

**Figure 11. Field effect black phosphorus devices.** (a) Optical image of a 12 nm thick black phosphorus field-effect device. (b) and (c) Conductance versus gate-voltage transport characteristics measured at 300K and 4K in the black phosphorus device shown in (a). (b) and (c) corresponds to channel lengths of 450 nm and 150 nm respectively.

| Channel length (nm) | Temp (K) | Mobility (e) (cm$^2$/Vs) | Mobility (h) (cm$^2$/Vs) | ON/OFF ratio |
|---|---|---|---|---|
| 450 | 300 | 12 | 35 | 600 |
| 450 | 4 | 14 | 31 | 40000 |
| 150 | 300 | 4 | 10 | 50 |
| 150 | 4 | 4 | 8 | 10000 |

**Table 3. Summary of the measured mobility and on/off ratio in black phosphorus field-effect devices.** We display mobility and on/off ratio measured for devices with two different channel lengths at two different temperatures.

**Conclusions**

We employed mechanical exfoliation to isolate few-layer black phosphorus flakes with thickness down to two single-layers. We found that the conventional mechanical exfoliation method, widely employed to fabricate graphene, typically produces low yield of atomically thin flakes. Therefore, we developed a modified version of the mechanical exfoliation method that employs a silicone-based transfer layer to optimize the yield of atomically thin black phosphorus flakes. General guidelines to identify and to determine the thickness of few-layer black phosphorus flakes using optical microscopy, Raman spectroscopy and transmission electron microscopy are presented. Transmission electron microscopy measurements demonstrate that freely-suspended black phosphorus flakes are stable and highly crystalline. The band structure of black phosphorus was calculated by density functional theory, finding a large thickness dependence of the bandgap. The exciton binding energy of black phosphorus with different number of layers was also calculated to estimate the optical bandgap, obtaining results consistent with preliminary photoluminescence experiments. We also studied the environmental stability of the fabricated flakes finding that air moisture is adsorbed on their surface due to a high hydrophilic character of few layer black phosphorus flakes. We observed that long term exposure to ambient conditions degrades thin black phosphorus flakes but few-layer flakes remain stable for several days. We showed that the aging of the black phosphorus flakes is slow enough to allow nanofabrication processing, presenting black phosphorus field-effect transistors with strong ambipolarity, mobilities of ~35 $cm^2$/Vs and current on/off ratios of ~600 at room temperature even without need of any post-fabrication cleaning step.

**Materials and methods**

*Sample fabrication and characterization*

Black phosphorus nanosheets have been prepared on PDMS-based substrates (Gel-Film® WF 6.0mil ×4 films) by mechanical exfoliation of commercially available black phosphorus (99.998%, Smart Elements) with blue Nitto tape (Nitto Denko Co., SPV 224P). We identify few-layer flakes under an optical microscope (Olympus BX 51 supplemented with an Canon

EOS 600D digital camera) and determined the number of layers by their opacity in transmission mode (see Supporting Information).

The topography of the flakes has been characterized by atomic force microscopy (Digital Instruments D3100 AFM) operated in the amplitude modulation mode

Raman spectroscopy was also performed (Renishaw *in via*) in a backscattering configuration excited with a visible laser light ($\lambda$ = 514 nm). Spectra were collected through a 100× objective and recorded with 1800 lines/mm grating providing the spectral resolution of ~ 1 cm$^{-1}$. To avoid laser-induced heating and ablation of the samples, all spectra were recorded at low power levels P ~ 500 µW and short integration times (~1 sec). Photoluminiscence measurements have been carried out with the same Renishaw *in via* setup but longer integration times were needed (~ 60-180 sec).

*Electron diffraction analysis*

Electron diffraction patterns were simulated using MacTempas software [48], in which five different scattering potential planes were stacked, one for each P plane and one empty one for the propagation between the P planes. Note that a monolayer of black phosphorous consists of two of such P planes. For the calculations of the diffraction patterns we have used *a*= 3.31, *b*=10.4, *c*=4.37, with space group *Cmca*, whereby the *b* axis is the stacking direction. For making the models of the monolayer etc we have divided the unit cell in 10 subslices of 0.104 nm thickness each, four of which contain P and the others are empty. With these subslices any stacking can be made for instance for a monolayer. The signature for every uneven number of layers is that the reflections with $h+l = 2n$ with $h$ and $l$ both uneven have a significant intensity. In case of a monolayer, the strongest of these reflections is the 101 reflection. We have taken the intensity of the 200 reflection to compare with the 101, because the intensity of the 002 reflection depends the atomic coordinates of the P atom, and we cannot exclude that these coordinates are different for a monolayer (See Supplementary information). The I(101)/I(200) ratio decreases for the series 1,3,5 etc layers as is to be expected.

*Field effect devices fabrication*

Few-layer black phosphorus transistors with Al/Ti (70 nm/5 nm) contacts are fabricated using standard electron-beam lithography and metal evaporation followed by a lift-off process.

*Crystal and electronic structure calculations*

The crystal structure has been computed with the SIESTA code [29] using its implementation for van der Waals interactions [30]. For the calculation of the electronic structure we use the CRYSTAL code [31] because of its implementation of the B3LYP functional [49]. We have decided to use a larger portion of HF exchange than other popular choices of hybrid functional, for instance the Heyd-Scuseria-Ernzerhof hybrid functional [50]. This functional includes less HF exchange and tends to underestimate the fundamental gap [51].

*Exciton binding energy calculation*

We employ the Wannier effective-mass theory [38] in two-dimensions, *i.e.*, we consider that for length scales of the exciton Bohr radius, the electron and hole are confined within the plane. Since the black phosphorus flakes are placed at the interface between two media with different dielectric constants (air and $SiO_2$) [39], we used the effective in-plane 2D interaction derived by Keldysh [52] to calculate the potential energy of the electron-hole pair (see Supporting Information). Moreover, due to the anisotropic effective masses for the electron and hole quasiparticles in the *x* and *y* directions, one cannot resort to a simplified isotropic calculation like the one carried out for transition metal dichalcogenides [39]. To calculate the binding energy we use a standard variational scheme with a 2D wave function that accounts for the material anisotropy (see Supporting Information for more details).

**ACKNOWLEDGMENT**

A.C-G. acknowledges financial support through the FP7-Marie Curie Project PIEF-GA-2011-300802 ('STRENGTHNANO'). This work was supported by the Dutch organization for Fundamental Research on Matter (FOM), the Spanish Ministry of Economy and Innovation through Grants Nos. FIS2010-21883-C02-2, FIS2012-37549-C05-03 and CONSOLIDER CSD2007-0010, the Generalitat Valenciana under Grant PROMETEO/2012/011, and the Ramón y Cajal Program (E.P). KLMA acknowledges support from the CCC of the Universidad Autónoma de Madrid.

**Supporting Information Available**: Supporting Information includes: Analysis of the optical transmittance, Raman spectra acquired on different flakes, laser induced damage in the black phosphorus flakes, calculation of the intensities of the diffraction spots and a detailed description of the exciton binding energy calculation.

Supporting Information:

# Isolation and characterization of few-layer black phosphorus

*Andres Castellanos-Gomez[1,]\*, Leonardo Vicarelli[1], Elsa Prada[2], Joshua O. Island[1], K. L. Narasimha-Acharya[2], Sofya Blanter[1], Dirk Groenendijk[1], Michele Buscema[1], Gary A. Steele[1], J. V. Alvarez[2], Henny W. Zandbergen[1], J. J. Palacios[2] and Herre S.J. van der Zant[1]*

[1] Kavli Institute of Nanoscience, Delft University of Technology, Lorentzweg 1, 2628 CJ Delft (The Netherlands).

[2] Departamento de Física de la Materia Condensada, Instituto de Ciencia de Materiales Nicolás Cabrera (INC), and Condensed Matter Physics Center (IFIMAC), Universidad Autónoma de Madrid, Cantoblanco, 28049 Madrid (Spain).

a.castellanosgomez@tudelft.nl

**Supporting Information content:**

1. **Comparison between conventional and PDMS-based mechanical exfoliation**

2. **Optical transmittance analysis**

3. **Thickness dependence of the Raman peak positions**

4. **Flake-to-flake variation of the Raman spectra**

5. **Simultaneous photoluminescence and Raman measurement**

6. **Laser induced damage in the black phosphorus flakes**

7. **Intensities of the simulated electron diffraction patterns**

8. **Details of the exciton binding energy calculation**

**Optical transmittance analysis**

Figure S1 presents low magnification optical images of samples fabricated by conventional mechanical exfoliation (employing blue Nitto adhesive tape) or using the PDMS-based exfoliation method described in the main text. While the conventional exfoliation method yields low density of black phosphorus flakes and leaves many traces of glue on the surface (marked by arrows in Figure S1b), the PDMS-based transfer method yield higher density of flakes, already visible in the low magnification images. Moreover, the sample remains clean and many of the transferred flakes are very thin (less than 10 nm thick), see inset in Figure S1c.

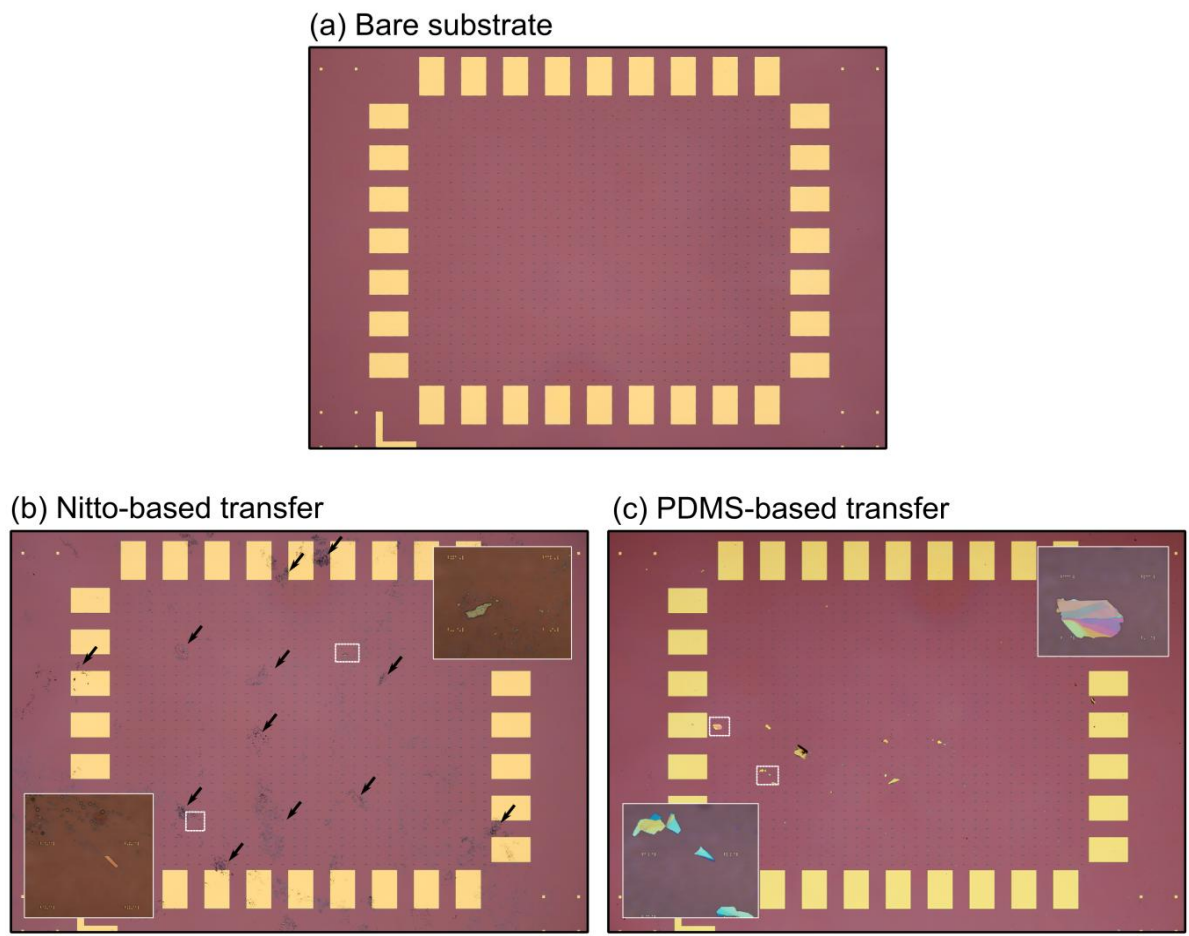

**Figure S1. Optical images of conventional and PDMS-based exfoliated samples.** Optical image of a bare SiO$_2$/Si substrate, pre-patterned with bonding pads and identification markers. (c) and (d) are optical images of two samples fabricated by conventional mechanical exfoliation (employing adhesive Nitto tape) and the PDMS-based mechanical exfoliation method described in the text. The insets in (a) and (b) shows a zoomed-in image of the regions marked with dotted white squares.

**Optical transmittance analysis**

Figure S2a and S2b show two false color maps of the transmittance of few-layer black phosphorus flakes. In order to quantitatively obtain the transmittance, an optical image is acquired using a digital camera attached to the trinocular of the microscope. The values of the red, green and blue channels are summed to make a matrix of intensities per pixel. Special attention is taken to avoid using images with some channel saturated or underexposed. Then the transmittance is obtained by dividing the matrix by the average intensity on the bare substrate. The resulting matrix can be plotted in a false color image as Figure S1a and S1b. Figure S1c presents the intensity histograms acquired in the boxed areas in panels (a) and (b). The histograms show peaks corresponding to the substrate (100% transmittance) and the different black phosphorus layers. The absorption of the different layers is a multiple of 2.8% and we thus infer that this value corresponds to the optical absorption of a single-layer black phosphorus.

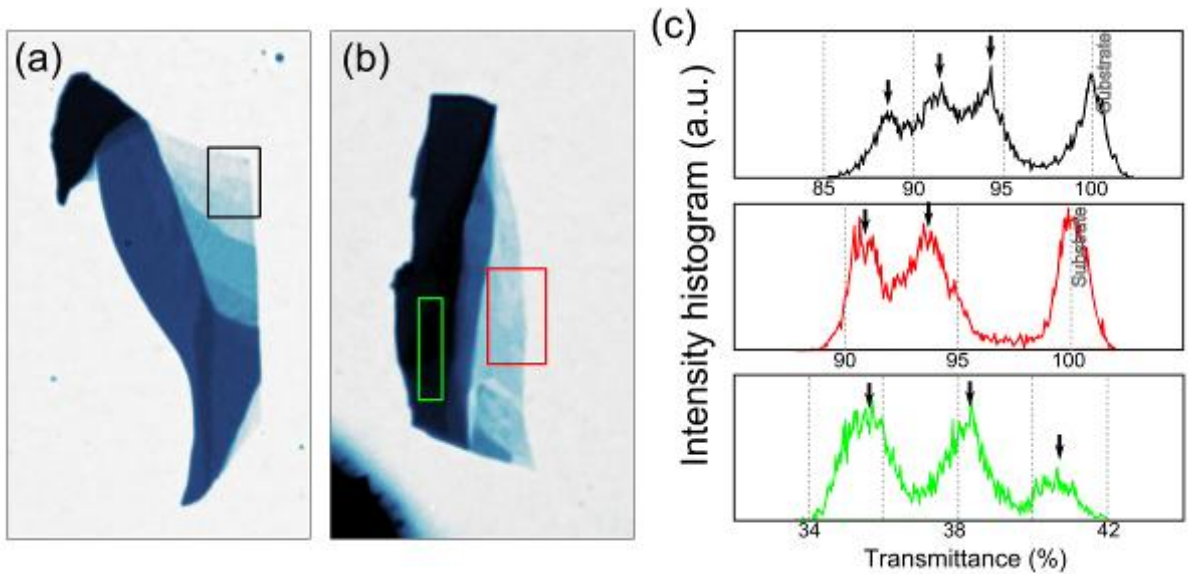

**Figure S2. Analysis of the optical transmittance.** (a) and (b) False color images of the optical transmittance of black phosphorus flakes deposited onto PDMS substrates. (c) Transmittance histograms acquired on the regions highlighted with rectangles in (a) and (b).

**Thickness dependence of the Raman peak positions**

Figure S3 shows the thickness dependence of the position of the Raman peaks associated to the lattice vibrations of the black phosphorus flakes. The Raman modes of the thinner flake (bilayer) moves towards higher Raman shift and thus, it can be used to distinguish bilayer flakes from thicker flakes.

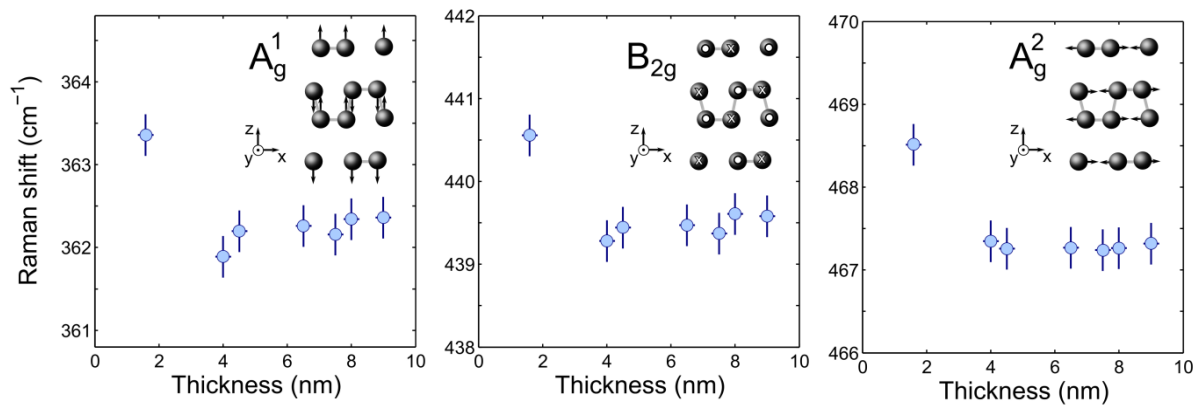

**Figure S3. Thickness dependence position of the Raman peaks.** The Raman shift of the $A^1_g$, $B_{2g}$ and $A^2_g$ modes has been determined for black phosphorus flakes with different thicknesses.

**Flake-to-flake variation of the Raman spectra**

We found that while the intensity ratio between the $A^1_g$ peak and the silicon peak depends monotonically with the number of layers (see Figure 4c of the main text), the ratio between the intensity of the $B_{2g}$ (or $A^2_g$) and the silicon peak varies from flake to flake. Figure S4 shows a comparison of Raman spectra acquired for flakes with similar thicknesses (as determined by AFM). The ratio between the $B^2_g$ and $A^2_g$ modes is inverted in the spectra measured in different flakes. This variation might be due to different crystalline orientations of the different flakes, as the $B_{2g}$ and $A^2_g$ are due to vibrations within the layer plane.

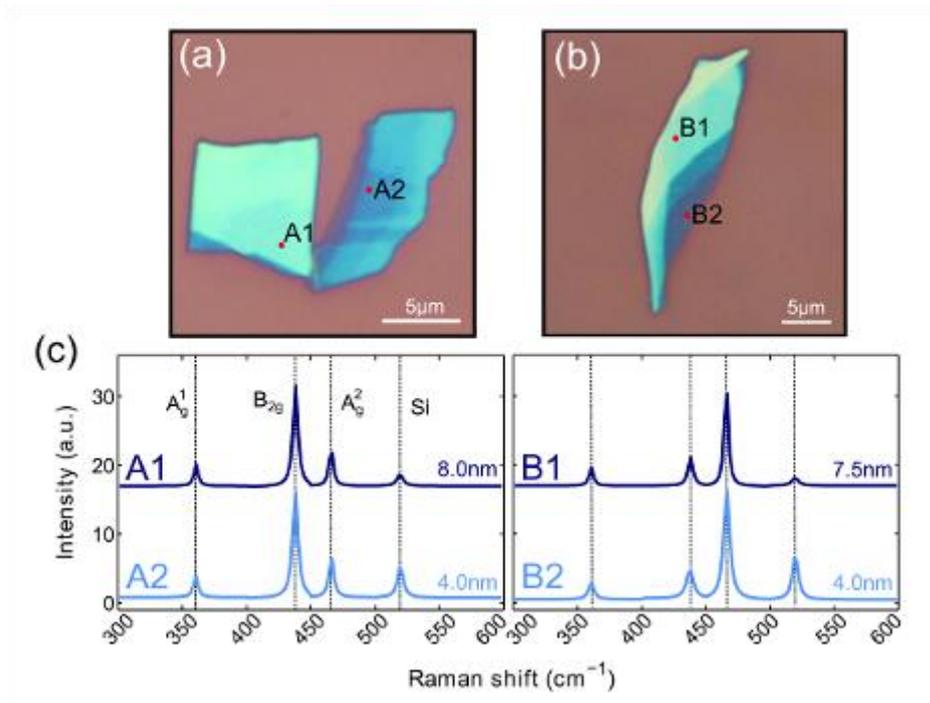

**Figure S4. Raman spectra of different black phosphorus flakes.** (a) and (b) Optical images of black phosphorus flakes deposited onto $SiO_2$/Si substrates. (c) and (d) show the Raman spectra acquired on the regions indicated with red points in (a) and (b).

**Simultaneous photoluminescence and Raman measurement**

Figure S5 shows the simultaneously acquired Raman and photoluminescence spectra of a bilayer black phosphorus flake. Note that the Raman and PL peaks are very weak in comparison with the silicon peak, probably due to the low optical absorption of the flake. The integration time was limited to 30 seconds due to the laser-induced damage (see next section). The intensity of the Raman peaks is smaller than the PL peak, similarly to what was observed for single layer $MoS_2$.

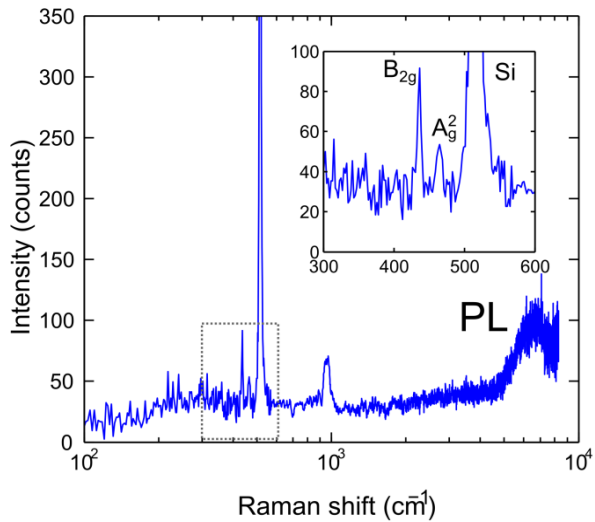

**Figure S5. Raman spectra of different black phosphorus flakes.** (a) and (b) Optical images of black phosphorus flakes deposited onto $SiO_2$/Si substrates. (c) and (d) show the Raman spectra acquired on the regions indicated with red points in (a) and (b).

**Laser induced damage in the black phosphorus flakes**

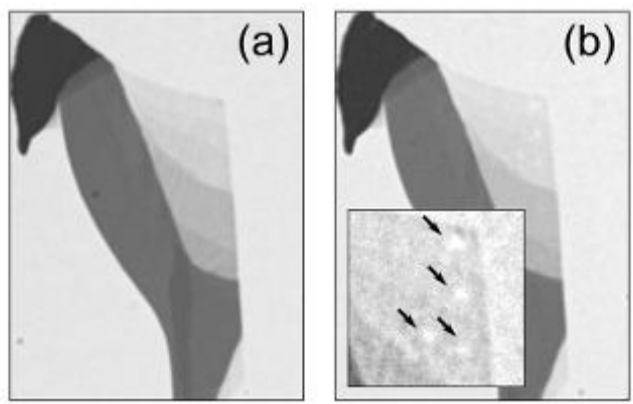

**Figure S6. Laser-induced damage in black phosphorus flakes.** (a) and (b) Optical transmittance images of a black phosphorus flake deposited onto a PDMS substrate before and after carrying out photoluminescence measurements. A zoom-in of the region damaged with the laser, indicated with arrows, have been inset in (b) .

As the thin black phosphorus flakes are damaged during the photoluminescence measurements the presence of the photoluminescence peak observed in bilayer samples might be due to laser-induced chemical modification of the black phosphorus. First we ruled out amorphization induced by the laser heating as the amorphous phosphorus emits at much lower energy (1.1-1.3 eV). We strongly irradiate a thick black phosphorus flake with a high power laser until achieving laser-ablation to test the changes induced in the photoluminescence spectra due to the chemical modification. Figure S7a and S7b show the optical images of a black phosphorus flake before and after the laser ablation process. The irradiated region develops a photoluminescence peak at 2.0-2.1 eV which we attribute to the laser induced chemical modification (e.g. oxidation). Therefore, the chemically modified phosphorus emits at much higher energy than that observed in bilayer black phosphorus.

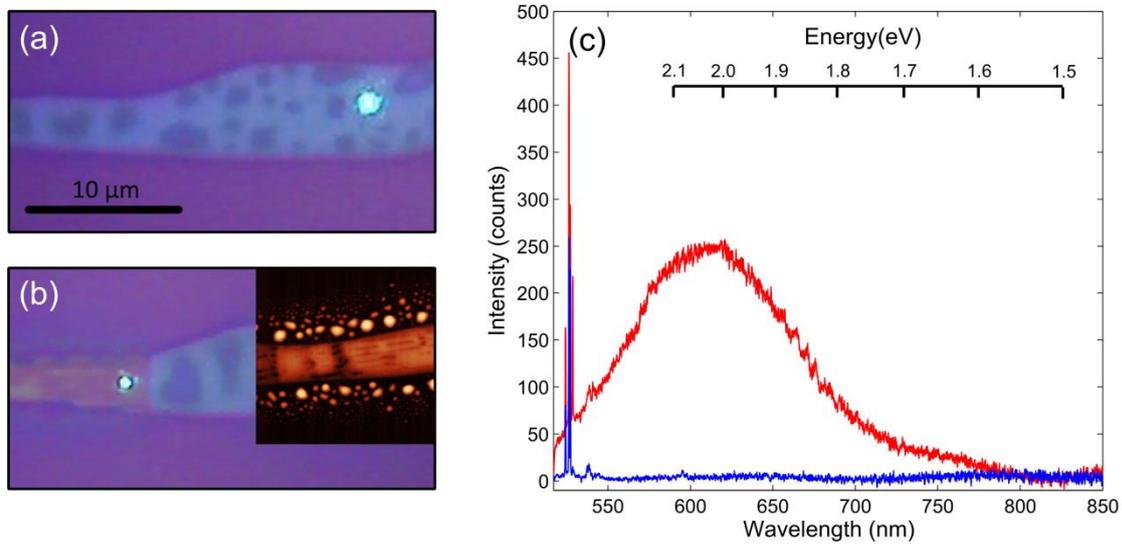

**Figure S7. Laser-induced chemical modification the black phosphorus flakes.** (a) and (b) Optical images of a thick black-phosphorus flake before and after scanning a high power laser (20 mW with a 400 µm diameter spot) over part of the flake. (inset in b) AFM topography image of the laser-ablated area. (c) Photoluminescence spectra acquired on the pristine flake (blue) and on the laser-modified flake (red).

**Intensities of the simulated electron diffraction patterns**

Table S1 shows the results of the calculation of the intensities of the diffraction spots performed using the MacTempasX software package, version 2.3.43, with a multi-slice approach [1].

| Z (P) | I(002)/I(200) | I(101)/I(200) | I(004)/I(002) |
|---|---|---|---|
| 0.08 | 0.62 | 2.56 | 0.07 |
| 0.085 | 0.5 | 2.46 | 0.14 |
| 0.09 | 0.4 | 2.36 | 0.23 |
| 0.095 | 0.29 | 2.27 | 0.42 |
| Exp data 1 | 0.40 | 0.29 | ~0.06 |
| Exp data 2 | 0.43 | 0.42 | ~0.06 |
| Exp data thick | 0.5 | 0.01 | 0.07 |

**Table S1.** Effect of the *z* position of the P atoms on the simulated intensity ratios of various reflections for a monolayer black phosphorus. The experimental values have been included for comparison.

**Details of the exciton binding energy calculation**

Since the thin black phosphorus samples are placed over a substrate ($SiO_2$), there is a large dielectric contrast between the vacuum on top of the monolayer and the weak dielectric below. This is similar to recent studies for TMDs, $MoS_2$ in particular. For this reason, to calculate the potential energy of the electron-hole pair separated by a distance $r = \sqrt{x^2 + y^2}$, we use the effective in-plane 2D interaction derived by Keldysh,

$$V_{2D}(r) = -\frac{e^2}{4(\varepsilon_1 + \varepsilon_2)\varepsilon_0} \frac{1}{r_0} \left[ H_0\left(\frac{r}{r_0}\right) - Y_0\left(\frac{r}{r_0}\right) \right].$$

This is the interaction felt by two charges living in a slab-geometry material of thickness d and dielectric constant $\varepsilon$, where $\varepsilon_1$ and $\varepsilon_2$ are the dielectric constants of the upper and lower media, all given in terms of the vacuum permittivity $\varepsilon_0$.

The functions $H_0$ and $Y_0$ are the Struve and the Bessel of the second kind ones, and $r_0 \equiv d\varepsilon/(\varepsilon_1 + \varepsilon_2)$ is the screening length signaling the polarizability of the three-region system. This interaction behaves as a $1/(\bar{\varepsilon} r)$ screened Coulomb potential (where $\bar{\varepsilon} = (\varepsilon_1 + \varepsilon_2)/2$) at long range and as a 2D logarithmic one at short range, where the crossover is determined by $r_0$.

Of course, a monolayer is somehow an extrapolation or limiting case of a thin slab material, but we will assume that the Keldysh interaction still constitutes a fair first approximation for a calculation based on an effective theory when $d$ is the interlayer distance. In our case $d = 5.49$ Å, $\varepsilon_1 = 1$ for vacuum, $\varepsilon_2 = 3.8$ for the SiO$_2$ substrate, and $\varepsilon = 10$ [2,3]. The screening length is thus $r_0 = 10.79$ Å.

As we saw previously, monolayer black phosphorus has pretty anisotropic masses for the electron and hole quasiparticles in the $x$ and $y$ directions, so that we cannot resort to a simplified isotropic calculation like the one carried out with TMDs [4]. This fact, in combination with the non-trivial form of the Keldysh interaction, prevents us to perform an exact calculation for the excitonic properties. Thus, to calculate the binding energy we use a standard variational scheme with a 2D wavefunction that accounts for the material anisotropy

$$\Psi(x,y) = \sqrt{\frac{2}{\pi \lambda_1 \lambda_2}} e^{-\sqrt{(x/\lambda_1)^2 + (y/\lambda_2)^2}},$$

where the two parameters $\lambda_1$ and $\lambda_2$ are varied independently. Note that this variational wavefunction is a very good approximation in the limit of weak screening, where $V(r) \to 1/r$ [5]. Actually, it is the exact ground state for the isotropic case $\lambda_1 = \lambda_2 \equiv \lambda$ in this limit, in which case $\lambda = a_0 \bar{\varepsilon} / 2\mu$, where $a_0 = 0.592$ Å is the Borh radius. For non-zero polarizability, the wavefunction is no longer exact but has the correct asymptotic behavior. In terms of the electron-hole reduced masses in the $x$ and $y$ directions, $\mu_x$ and $\mu_y$, the kinetic energy can be found analytically yielding

$$T(\lambda_1, \lambda_2) = \frac{\hbar^2}{4}\left(\frac{1}{\mu_x \lambda_1^2 + \mu_y \lambda_2^2}\right).$$

The potential energy $V(\lambda_1, \lambda_2)$, though, has to be found by numerical integration. The exciton binding energy $E_b$ is then found by minimizing the total energy $T(\lambda_1 + \lambda_2) + V(\lambda_1 + \lambda_2)$, where the optimum values of $\lambda_1$ and $\lambda_2$ are the exciton radii in the *x* and *y* directions. The result of this procedure for different number of layers in shown in Table 1 of the main text.

**Supporting Information references**

[1] Kilaas R. In: Bailey GW, editor. 45th annual proceedings of EMSA. Baltimore (MD): San Francisco Press; 1987. p. 66.

[2] H. Asahina and A. Morita, J. Phys. C: Solid State Phys. **17**, 1839 (1984).

[3] Here, and following Landau, we use a geometric mean for the dielectric constant $\varepsilon = (\varepsilon_x \varepsilon_y \varepsilon_z)^{1/3}$, where for bulk b-P we have $\varepsilon_x = 12.5$, $\varepsilon_y = 10$ and $\varepsilon_z = 8$, according to [2]. Note that, although anisotropic, the dielectric constants are rather similar in the three directions.

[4] T. C. Berkelbach, M. S. Hybertsen, and D. R. Reichman, Phys. Rev. B **88**, 045318 (2013).

[5] A. Schindllmayr, Eur. J. Phys. **18**, 374 (1997)